\documentclass[11pt,twoside]{article}

\usepackage{asp2006}
\usepackage{epsf}
\usepackage{psfig}
\usepackage{lscape}

\begin{document}
\title{C- and O-Rich Miras and Galactic Structure}   
\author{Michael Feast}   
\affil{Astronomy Department, University of Cape Town\\
mwf@artemisia.ast.uct.ac.za}    

\begin{abstract} 
This paper summarizes the conclusions of an extensive investigation
of variable carbon AGB stars in our Galaxy. The zero-point of the 
period- $M_{bol}$ relation for Galactic C-Miras is found to be close
to that in the LMC. The mean age of Galactic C-Miras is $\sim$ 1.8 Gyr
and their initial masses $\sim 1.8 M_{\odot}$ with some evidence that
age decreases and initial mass increases with increasing 
pulsation period.
Kinematic studies of Galactic C- and O-Miras show that the relative
frequency of these two Miras classes depends on the age of the
parent population. The lack of carbon stars in the Galactic Bulge
seems to be due to a high 
oxygen abundance. The 
velocity dispersion of O-Miras allows one to distinguish between
two possible models  of Galactic kinematics.

\end{abstract}


\section{Introduction}
The kinematics of the large amplitude, oxygen-rich AGB variables 
known as
O-Miras have long been known to give valuable clues to the ages and
evolution of AGB stars as well as being important for Galactic
structure studies (e.g. Feast 1963; Feast \& Whitelock 2000). Until
recently the Galactic carbon-rich Miras (C-Miras) have been much less
intensively studied. The present paper summarizes the results of 
a kinematic and photometric study of Galactic
C-Miras leading to a determination of their absolute magnitudes and
ages. It then goes on to discuss what C- and O-Miras tell us about
Galactic structure and composition. The paper is based on three
recently published papers (Whitelock et al. 2006; 
Menzies, Feast, \& Whitelock 2006;
Feast, Whitelock, \& Menzies 2006).

It has been known for a long time that an optically selected sample
of C-Miras in the LMC shows a well defined $M_{bol}$ - log\,P relation
(Feast et al. 1989). It is now possible to extend this relation to
longer periods using data for obscured C-Miras (Whitelock et al. 2003).
The relation which extends from about 160 to 1000 days has the form
\begin{equation}
M_{bol} = -2.54 \log\,P + 1.87
\end{equation}
where the distance modulus of the LMC has been taken to be 18.50mag.
The scatter about this relation is only 0.17mag. One aim of
the present work was to derive the zero-point of this relation
for Galactic C-Miras.
\section{Galactic C-Miras: Absolute Magnitudes, Velocity Dispersions 
and Ages}
Multi-epoch JHKL observations were obtained of 239 C-rich variables
allowing classification and period determination. Mean bolometric
magnitudes were obtained by combining these data with IRAS or
MSX observations (details in Whitelock et al. 2006). Combining these 
results
with data in the literature and some new optical radial velocities
we have available for analysis 177 C-Miras with known bolometric
magnitudes
and 
radial velocities from optical or 
millimetre (CO)  observations (details in
Menzies et al. 2006). Assuming a PL relation of the form given in 
eq. 1, we found clear evidence of differential galactic rotation. 
The zero-point of this relation was then adjusted until,
using a standard formulation of differential rotation, we recovered
the Oort constant $A$ derived from Cepheid proper motions
(Feast \& Whitelock 1997). In this way we obtained a Galactic
zero-point of $+2.06 \pm 0.24$, not significantly different from
that for LMC C-Miras (eq.1). In the kinematic analysis, we used 
only stars within
$\rm \pm 2 kpc$ of the solar distance from the Galactic Centre. 
The data are not sufficient to allow us
to derive an independent PL slope.
\begin{table}[!ht]
\caption{Velocity dispersions of C-Miras}
\smallskip
\begin{center}
{\small
\begin{tabular}{cccccc}
\tableline
\noalign{\smallskip}
Group & Period & $\alpha$ & Age & Mass & N\\
      & [days] & [$\rm km\,s^{-1}$] & [Gyr] & $\rm M_{\odot}$ & \\
\noalign{\smallskip}
\tableline
\noalign{\smallskip}
All & 521 & $27 \pm 2$ & $ 1.8 \pm 0.4$ & $1.8 \pm 0.2$  & 149\\
1   & 373 & $32 \pm 3$ & $ 3.1 \pm 0.9$ & $1.5 \pm 0.2$  &  49\\
2   & 521 & $24 \pm 3$ & $ 1.3 \pm 0.5$ & $2.1 \pm 0.3$  &  50\\
3   & 655 & $24 \pm 2$ & $ 1.3 \pm 0.3$ & $2.1 \pm 0.2$  &  50\\
\noalign{\smallskip}
\tableline
\end{tabular}
}
\end{center}
\end{table}

The residual velocities from differential rotation were then used
to derive velocity dispersions. Table 1 shows
$\alpha$, the velocity dispersion
radial from the Galactic Centre, for the whole sample and the sample 
divided by period,
as well as the numbers (N) of stars involved. The ages and initial 
masses were derived using
the relation between the kinematics and age of local dwarfs (Nordstr\"om
et al. 2004) based on Padova models (Girardi et al. 2000). There is 
some evidence
of a trend of age with period, the longer period stars being
younger. This is qualitatively similar to the trend of age with period
shown by O-Miras (e.g. Feast \& Whitelock 2000). The results 
for the C-Miras
contrast with the younger ages and higher initial masses
sometimes suggested for the bulk of C-Miras.
 
\section{The Relative Frequency of Cool Carbon- and Oxygen-Rich Stars}
  In recent years there has been a considerable amount of work
based on the hypothesis that the relative frequency of
carbon to oxygen rich (M-type) cool stars is a simple measure of
the overall metallicity 
(normally characterized by [Fe/H]) of the stellar system 
or part of the system
in which they are found. The results of the last section allow
one to test this in the limited case of the C- and O-Miras and to
extend the discussion to cool stars generally.
\begin{table}[!ht]
\caption{Velocity Dispersions of O-Miras}
\smallskip
\begin{center}
{\small
\begin{tabular}{ccc}
\tableline
\noalign{\smallskip}
Period & $\alpha$ & N\\
 $[\rm days]$ &[$\rm km\,s^{-1}$] & \\
\noalign{\smallskip}
\tableline
\noalign{\smallskip}
175 & $ 76 \pm 8 $ & 17 \\
228 & $ 69 \pm 6 $ & 24 \\
272 & $ 46 \pm 4 $ & 26 \\
324 & $ 50 \pm 4 $ & 40 \\
383 & $ 45 \pm 4 $ & 32 \\
453 & $ 28 \pm 4 $ & 15 \\
\noalign{\smallskip}
\tableline
\end{tabular}
}
\end{center}
\end{table}

Table 2 shows the velocity dispersion radial from the Galactic Centre
for local O-Miras as a function of period. The data are derived
from Feast \&  Whitelock (2000) and show the dependence of
$\alpha$ (and hence age) on period for these stars. 
A comparison of Tables 1 and 2 shows
that there appears to be no significant populations of C-Miras
with $\alpha$ in the range 40 to $\rm 80\, km\,s^{-1}$ whilst there are
strong O-Mira populations in this range. Since age increases
with $\alpha$, this implies that there are O-Miras in an (old) age 
range where there is not a significant C-Mira population. Thus
the ratio of C- to O-Miras will depend on the age distribution
of any population considered and not simply on [Fe/H]. This result
may be seen in a somewhat different way for Miras in the solar
neighbourhood. For an optically selected sample Wood \& Cahn (1977)
found that the ratio of the number of O-Miras to C-Miras was about
14. However, in a local sample of dust enshrouded Miras which 
is biased to longer period (and hence younger) stars this
ratio is about unity (Olivier, Whitelock, \& Marang 2001).

It is useful to consider in this connection the stellar population of
the Galactic Bulge. With one possible exception, this contains 
no C-Miras or any normal carbon stars. However, it does contain
O-Miras with a wide range of periods. These range up to 700 days
in the SgrI window (Glass et al. 1995) and to even longer periods
near the Centre (Wood, Habing, \&  McGregor 1998). It has long been known
that this implies a considerable age range in the Bulge. The age range
implied overlaps with that expected for C-Miras (table 1). This
suggests that the absence of C-Miras from the Bulge is not
simply an age effect.

In fact, in the past it was usually assumed that the absence of
carbon stars in general from the Bulge was due to a high metal
abundance. It is, however, now known that Bulge K giants have
a mean [Fe/H] of $-0.10$ and M giants $-0.19$ (Fullbright, McWilliam,
\& Rich 2005;
Rich \& Origlia 2005). These values are close to the mean value for the
local thin disc, [Fe/H] = $-0.14$ (Nordstr\"om et al. 2004). Thus neither
age nor metallicity can apparently explain the absence of carbon stars
from the Bulge. However, it has been know for some while (Glass et al. 1995)
that the JHK colours of O-Miras in the Bulge differed from those in
the LMC and this could be explained by a high oxygen abundance
(Feast 1996). Recent work (Rich \& Origlia 2005) now shows
that in Bulge M giants oxygen is overabundant with respect to 
comparable
local stars ($ \rm [O/Fe] \sim +0.3$) whilst several workers find
$\rm [O/Fe] \leq 0$ in 
relevant LMC populations. This suggests that 
oxygen abundance
may be a controlling factor in
the ratio of carbon to oxygen giants in stellar systems.

The above does not rule out the possible dependence of the relative
frequency of carbon stars on [Fe/H] at a given [O/Fe]. 
If that were the case we would expect that
in a system with a
spread in [Fe/H] the carbon stars would evolve preferentially
from the lower metallicity stars
([O/Fe] assumed constant). However, in the Galactic disc
the available work on carbon stars ( e.g. Abia et al. 2001)
suggests that their [Fe/H] is near 
that of the mean of local disc stars and 
not skewed to lower abundances. But further work on this is desirable.

\section{O-Miras and the Kinematics of the Galactic Disc}
 Two alternative interpretations
of the kinematics of local disc stars
are currently under discussion. Nordstr\"om et al. (2004)
found that the velocity dispersion of these stars increases steadily
with age from 1 to 10 Gyr (see their fig 31 where their $\sigma_{U}$
corresponds to the $\alpha$ used in the present paper). On the other hand
Quillen \& Garnett(2000, 2001) (see also Freeman
\& Bland-Hawthorn  fig 5
where their $U$ corresponds to the $\alpha$ of the present paper) 
suggest that the velocity
dispersion remains constant from 2 to 10 Gyr at about
$\rm 35 km\,s^{-1}$  and then jumps
suddenly to about $\rm 60 km\,s^{-1}$, the two regimes 
constituting the thin and thick
discs. This latter interpretation apparently leaves no room
for Galactic components with $\alpha$ in the range 35 to 
$\rm 60\,km\,s^{-1}$
where, as Table 2 shows there is a significant O-Mira population.
Since O-Miras are rare objects we expect them to be tracers of a
large main sequence population. Such a population is accounted for
naturally in the Nordstr\"om et al. interpretation of Galactic
kinematics. It is, however, difficult to see how it could be 
accommodated in the Quillen and Garnett model.

\section{Summary}

1. Assuming the $M_{bol}$ - period relation for Galactic C-Miras
 has the same slope as for those in the LMC, the zero-points in the
two systems  are the same within the uncertainties\\
2. Galactic C-Miras have ages of $\sim 1.8$ Gyr and initial masses
of $\sim 1.8 M_{\odot}$.\\
3. There is some evidence that the ages of C-Miras decrease with 
increasing period.\\
4. The ratio of C- to O-Miras in a population depends on the
age distribution of that population.\\
5. The lack of carbon stars in the Galactic Bulge is connected
to an oxygen overabundance.\\
6. The Galactic O-Miras favour the
interpretation of Galactic kinematics of Nordstr\"om et al.


\acknowledgements 
This paper is based on three papers mentioned in the text and I am
grateful to my colleagues, Patricia Whitelock, John Menzies, 
Freddy Marang and  M.A.T. Groenewegen for their contributions
to this work. In particular, I am especially indebted to
Patricia Whitelock and John Menzies with whom the 
data were analysed and the conclusions formulated.


\end{document}